\documentclass[12pt,preprint]{aastex}
\usepackage{graphicx}

\newcommand{\be}{\begin{eqnarray}}
\newcommand{\ee}{\end{eqnarray}}

\newcommand {\nbodypp}{\textsc{\mbox{nbody6\raise.4ex\hbox{\tiny++}}}}
\newcommand {\COri} {\mbox{$\theta^1{\rm{C}}\:{\rm{Ori}}$}}

\newcommand {\Msun} {\mbox{M$_{\odot}$}}

\newcommand {\SAML} {specific angular momentum loss}

\begin{document}

\title{Cluster-assisted accretion for massive stars}
\author{S. Pfalzner} 
\affil{I. Physikalisches Institut, University of Cologne, Z\"ulpicher Str. 77,
50937 K\"oln, Germany}
\email{pfalzner@ph1.uni-koeln.de}

\begin{abstract}
Gravitational interactions in  very young high-density stellar clusters can to some degree 
change the angular momentum in the circumstellar discs surrounding initially the majority of stars. 
However, for most stars the cluster environment alters the angular momentum 
only slightly. For example, in simulations of the Orion Nebula cluster (ONC) encounters reduce 
the angular momentum of the discs on average at most by 3-5\% and in the higher 
density region of the Trapezium 
by 15-20\% - still  a minor loss process. 
However, in this paper it is demonstrated that the situation is very different if one considers 
high-mass stars (M$^*>$10 M$_\sun$) only. 
Assuming an age of 2 Myr for the ONC, their discs have on average a  50-90\% lower angular momentum 
than primordially. This enormous loss in angular momentum in the disc should 
result in an equivalent increase in accretion, implying that the cluster 
environment boosts accretion for high-mass stars, thus
making them even more massive. 
\end{abstract}

\keywords{(stars:) circumstellar matter, (stars:) planetary systems: protoplanetary disks,  
(galaxies: star clusters), accretion, accretion disks}
\maketitle

\section{Introduction}

Observational evidence is mounting that most, if not all, stars are 
initially surrounded by discs. For example, \citet{lada:aj00} found that 
80-85\% of all stars in the Orion Nebula Cluster (ONC) possess a disc. 
In most cases young stars are not isolated but part of 
a cluster. This immediately poses the question of the relevance of interactions between
the cluster members. Back of the envelop estimates indicate that it is very unlikely 
that these cluster stars themselves collide, and, although the cross-section for   
star-disc systems is a lot larger (disc size $\sim$ few 100 AU), gravitational interactions 
between such star-disc systems should only play a minor role. This view is supported by
simulations by \citet{scally:mnras01}. However, initial substructuring or gas in the
cluster could change the dynamics that encounters could play a role in the early
stages \citep{scally:mnras02,bonnell:98}.

As already noted by \citet{mestel:qjras65} and \citet{spitzer:proc68},  the temporal development
of the angular momentum in the disc is of vital importance for our understanding of the late 
stages of star formation. The typical observed angular momentum of the cloud cores 
from which the stars develop is about three orders of magnitude larger than the maximum 
that can be contained in a single star \citep{bodenheimer:araa95}. For the formation of stars 
and planetary systems it is essential that angular momentum is transported outwards from the 
inner regions of the disc in some way, so that accretion is at all possible. Many different 
processes have been suggested for this angular momentum transport
\citep[see review by ][]{larson:mnras02}. 
One possible mechanism for accretion-enabling angular momentum transport is 
gravitational interaction. However, it was demonstrated in \cite{pfalzner:aa07} that 
in the ONC the specific angular momentum loss (AML) averaged over all stars 
(low-mass to massive) 
is at most 3-5\% increasing to 15-20\% in the Trapezium region (the region within 0.3 pc from the
cluster center). 
So gravitational interactions of single stars seem, at first sight, to be of minor 
importance for accretion (although possibly not for planet formation). 

However, recent detailed simulations \citep{olczak:apj06} of the effect of 
encounters showed that although the average disc mass loss does not exceed 10--15\% of the 
disc mass even in the the Trapezium region, the situation is altered remarkably if one 
considers the massive stars alone. 
Due to mass segregation situated close to the cluster center, they function as gravitational foci 
for the other cluster stars resulting in a $\sim$ 60 - 100\% relative disc mass loss for 
massive stars \citep{pfalzner:aa06} in the first 2 Myr of the cluster development.

This raises the question of whether the same can be said for the AML
of the massive stars.  In the following it will be shown that indeed the average 
relative AML is extraordinarily higher for massive stars than lower mass stars.

\section{Cluster and encounter simulations}

As in \citet{pfalzner:aa07} the ONC was chosen as model cluster because 
its high density suggests that stellar encounters might be relevant for the evolution 
of circumstellar discs. In addition, it is one of the best-studied regions 
in our galaxy, so that observational constraints significantly reduce the range of modelling 
parameters. 
Combining results of the dynamics of the ONC with those of isolated 
star-disc encounters, allows 
the AML of the discs to be determined. The procedure is described in detail 
in \cite{pfalzner:aa07}, here only a short summary of the simulation method
is given since the main purpose of this investigation is to reanalyse the data 
with special emphasis on massive stars. 

The dynamical model of the ONC contains only stellar components ($\sim$ 4000 stars, 
with $\sim$ 700 of them in the Trapezium region) neglecting 
gas and the potential of the background molecular cloud OMC~1.
Cluster models were set up with a spherical density distribution $\rho(r)\propto r^{-2}$ 
and a Maxwell-Boltzmann velocity distribution. Some observations indicate a 
uniform density core of approximately 0.032pc radius \citep{mcc:02}, but our simulation results are not sensitive to this
difference in initial conditions. The masses were generated randomly according 
to the mass function given by \cite{kroupa:mnras93} in a range $50 \Msun \ge M^* \ge 0.08 \Msun$, 
apart from \COri, which was directly assigned a mass of 50 \Msun \ and placed at the cluster centre. 
The ONC was simulated for 13\,Myr -- the assumed lifetime of \COri.

The cluster simulations were performed with \nbodypp\ \citep{spurzem:mnras02} and 
the quality of the dynamical models determined by comparing them to the observational data  
at the approximate age of the ONC (1-2Myr). The quantities of 
interest were: number of stars, half-mass radius, number densities, velocity dispersion 
and projected density profile.
Most investigations were performed for the ONC in virial equilibrium. 
Information concerning all perturbing events of each stellar 
disc was recorded in an encounter list. 
It was assumed that only two-body encounters occur and 
that higher-order encounters are negligible, so that the effect of an 
encounter were investigated by considering it to be isolated from the rest of the cluster.

Earlier angular momentum transport investigations of star-disc encounters 
\citep{ostriker:apj94,hall:mnras96,pfalzner:aa05} were extended in \cite{pfalzner:aa07} 
to cover the parameter range necessary for modelling the ONC.
The disc surrounding the star was assumed to extend to $r_d=100$\,AU and the surface density 
to have a $1/r$-dependence initially. For a star of mass $M_1^*=1\,\Msun$
the angular momentum loss $\Delta J/J$ induced by the fly-by of a star of mass $M_2^*$ was determined
for  parabolic, prograde, coplanar encounters for low mass discs
and fitted by
%
\begin{eqnarray}
    \frac{\Delta J}{J} =   
    1.02 \left(\frac{M_2^*}{M_1^*+M_2^*}\right)^{0.5r_p}
    \exp \left[ - \sqrt {\frac {M_1^*\left(r-0.7r_p^{0.5}\right)^{3}} {M_2^*} }\right]
 \label{eqn:Fit_total}
\end{eqnarray}

where is $r_p$ is the periastron in units of the disc radius.
The low-mass assumption matches the observational evidence in the ONC that for most 
disc masses $m_d/M^* \ll$ 0.1. This reduces the complexity  
since low-mass discs do not significantly influence the encounter 
orbit, self-gravitation and viscosity can be neglected, and the results are scalable 
to other star masses. Considering  only parabolic coplanar, prograde encounters means
 that the results can only be interpreted as upper limits
\citep{heller:apj95,hall:mnras96,pfalzner:aa05}. 


The wide spectrum of star masses in the cluster requires the simulation results for 
$M_1^*=1\,\Msun$ to be generalized. Here the disc radius is scaled with the stellar mass
according to $ r_d =r_d(1\,\Msun)\sqrt{M_1^*[\Msun]}$. This seems intuitively right, but 
observational results are ambiguous: Although \cite{vicente:05} see a 
correlation between disc diameters and stellar masses using a sample of proplyds from 
\cite{luhman:apj00}, they detect no dependence in the data from \cite{hillenbrand:aj97}. 

The relative AML $\Delta J/J$
given by Eq.~\ref{eqn:Fit_total} 
is always larger than the relative mass loss $\Delta m/m$ and effects the disc at much 
larger distances. 
Given a sufficiently large perturber mass, even encounters
at $r_p = 20 r_d$ as distant as 20 times the disc radius 
can reduce the angular momentum in the disc by $\leq$ 10\% 
or more without disc mass loss.

\section{Results}

Combining the cluster dynamics with the encounter results, in the following the
dependence on the stellar mass of the encounter-induced AML of the discs in the ONC is investigated.
For the Trapezium region as well as the entire ONC, the AML  
(according to Eq.1 in \cite{pfalzner:aa07}) is larger than the mass loss for all stellar 
masses $M_1^*$. More importantly, both increase considerably for massive stars.
The largest difference between the mass and angular momentum loss is found in the
mass range of 5-15\Msun.
As the high-mass stars are also the ones that have the highest mass loss, this 
could simply reflect that this lost mass carries with it a high angular 
momentum. 
However, what one is really interested in is whether the specific
angular momentum in the remaining disc is lowered, as only 
this can eventually facilitate accretion of matter on to the star. The specific angular momentum 
$(\Delta J/J)_{mass}$ is 
defined by the ratio of the relative angular momentum and the relative disc mass  $\Delta m/m$ 
of the remaining disc
\be
\left(\frac{\Delta J}{J}\right)_{mass} := 
\left(1-\displaystyle {\frac{\Delta J}{J}}\right)/
\left(1-\displaystyle {\frac{\Delta m}{m}}\right).
\label{eq:j_total}
\ee
From here onwards it is always this specific angular momentum that is considered.
\cite{pfalzner:aa07} showed that for a single encounter there exists an upper limit of
60\% AML in the remaining disc. 
However, as \citet{pfalzner:apj04} and \citet{moeckel:06} showed, consecutive encounters can lead to a 
significant angular momentum transport. 

\begin{figure}
\resizebox{\hsize}{!}{\includegraphics[angle=-90]{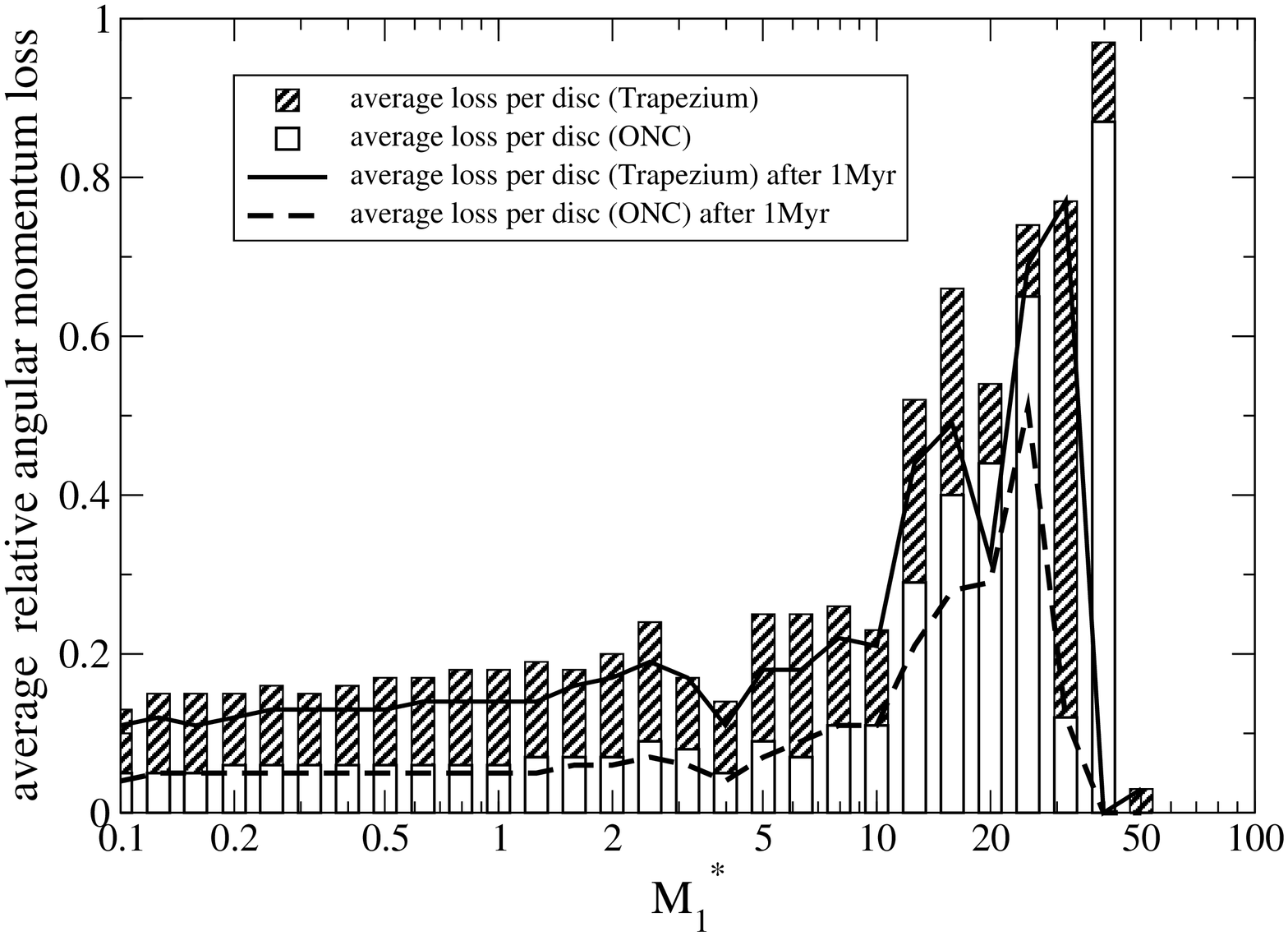}}
\caption{Relative SAML per disc mass as function of the mass 
$M_1$ averaged over 20 simulations for the Trapezium and the ONC. The bars show the 
result at 2 Myr, the lines the same at 1 Myr.}
\label{fig:angmassmean}
\end{figure}

So when determining the SAML as defined in Eq.~\ref{eq:j_total}, 
it rises from about 3--8\% for low-mass stars
to $\sim$ 40-70\% for high-mass stars in the ONC and from 15\%--20\% to $\sim$ 60-95\% in the Trapezium 
region (see Fig.~\ref{fig:angmassmean}).

\begin{figure}
\resizebox{\hsize}{!}{\includegraphics[angle=-90]{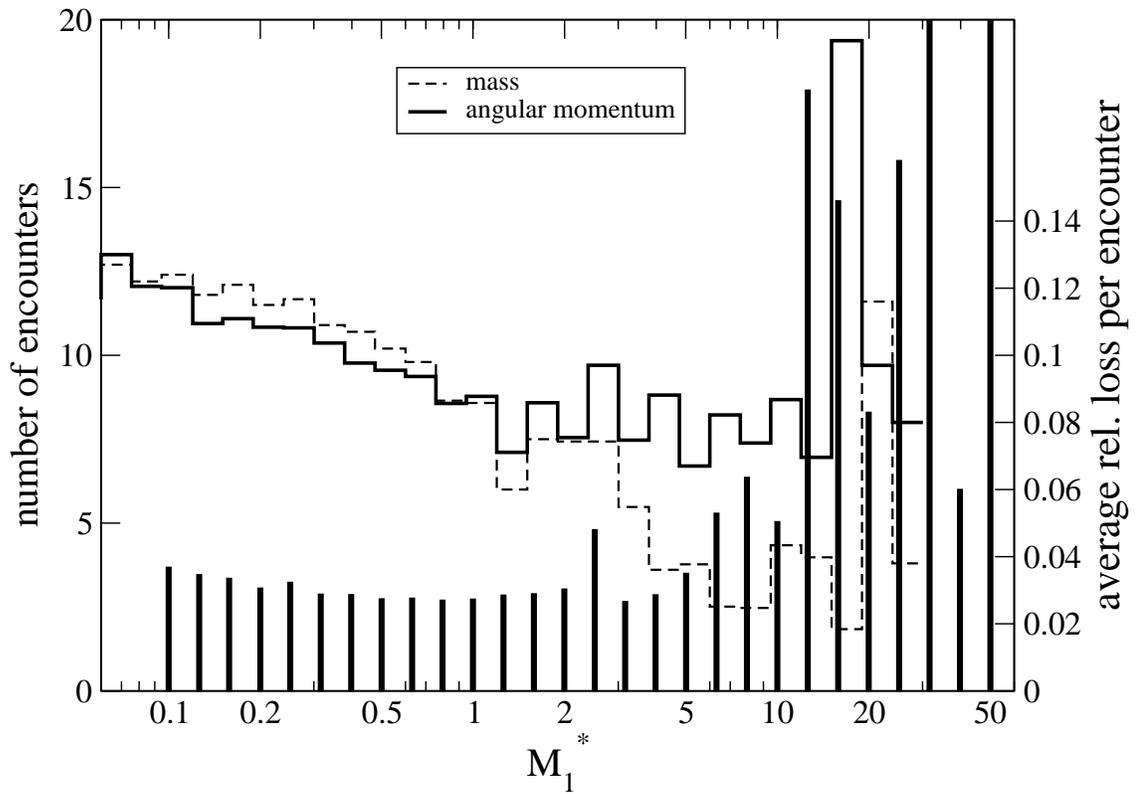}}
\caption{Relative average SAML per encounter (drawn line) and average 
number of encounters per star (black bars) as function of the mass $M_1$ averaged over 
20 simulations at 2 Myr. For comparison the relative average mass 
loss per encounter is shown as well (dashed line).}
\label{fig:angmassenc}
\end{figure}

How can a reduction to 5\% of the initial angular momentum occur? 
Clearly not in a single encounter, where only up to 60\% can be 
lost at a time (see \citet{pfalzner:aa07}). Fig.~\ref{fig:angmassenc} demonstrates  that the 
massive stars experience on average a much larger number of encounters than 
lower mass stars and the AML in  a single encounter is not as sensitive to the stellar mass 
as the mass loss is. The reason for the higher number of encounters is that massive stars act as 
gravitational foci for lower mass stars, which eventually leads to more pronounced SAML in 
the disc of massive stars.  


As the age of the ONC is not precisely known, with most estimates in the range of 1-2Myr, 
the question is how  sensitive  is this result 
to the ONC age?
Fig.~\ref{fig:angmassmean} also shows the SAML for the
case of a 1Myr old ONC. It can be seen that the overall result of a much higher SAML for massive 
stars still holds. 

The scaling of the disc size with the mass, although intuitively right, 
is not really observationally proven \citep{hillenbrand:aj97,luhman:apj00,vicente:05}.
Comparing above results to those obtained assuming an equal constant disc size (150AU) 
for all stars, we find that the now larger disc 
size for the low-mass stars leads to an increase in the SAML there, whereas the smaller size
for massive stars has the opposite effect. However, these are minor changes, leaving the overall
result of the much higher SAML for massive stars untouched.

Recent observations \citep[see][and references therein]{zhang:05} suggest that massive 
stars can be surrounded by high-mass discs $m_d > 0.1 M^*$.
Since the interaction dynamics involving high-mass discs is still poorly understood, 
we resolve to the crude method of setting $m_{disc}= M_1^*$ for all stars with 
$ M_1^*>$ 5 $M_\sun$  and simply add their disc mass to the stellar mass when determining
the AML using Eq.1 from \cite{pfalzner:aa07}. This leads to 
a $\sim$ 5-12\% reduction of the relative SAML, 
but again the general trend that massive stars lose more angular momentum than lower 
mass stars is unaffected. 

Up to now the ONC was treated as being in virial equilibrium. If the cluster is
allowed to expand ($Q_{vir}$=1) we find that the SAML is larger for all masses with the 
largest differences in the Trapezium region.
The additional SAML happens early on in the cluster development, where the density in the 
expanding cluster is initially higher in the center.

\section{Discussion and Conclusions}
\label{sec:discussion}

In this paper the dependence of the encounter-induced angular momentum 
loss in circumstellar discs has been investigated for the example of the ONC. 
This has been done 
by combining simulations of the cluster dynamics with investigations of SAML 
in isolated encounter events.

The main outcome is that although most stars experience only an average \SAML\
of 3-5\%, massive stars suffer much higher losses ($\sim$ 70-95\%).
The reason is that the massive stars act as gravitational foci, experiencing many 
more encounters than lower mass stars. Although the actual percentages might vary, the general 
trend of much higher SAML for massive stars is not very sensitive to modelling parameters
like disc size scaling with stellar mass, age of the ONC or virial coefficient of the
cluster.

The SAML was obtained assuming parabolic prograde, coplanar encounters, which as such can 
only be regarded as upper limits. Since stars of all masses are equally 
effected by this, the actual percentages might be somewhat lower, but it will still hold 
that massive stars have a much higher SAML than lower mass stars.

What are the consequences of such higher SAML for massive stars?
The SAML is connected to  many particles moving on highly eccentric orbits around the 
central star after the encounter. This does not in itself ease accretion straight 
away. However, viscosity-driven processes may cause 
the angular momentum to be redistributed in the disc, eventually easing accretion.
The correlation between angular momentum and accretion means that
the enormous loss in disc angular momentum for massive stars induced by the cluster environment
should result in an equivalent increase in accretion. 
This happens predominantly in the 
cluster center, but as well to a lesser degree further out. 
The interaction with the other cluster members thus triggers the accretion in the massive stars, 
a process which could be described as {\it cluster-assisted accretion}.

In Section 2 the SAML was deduced assuming that
only one of the stars is surrounded by a disc - contradicting the initial configuration where 
all stars are initially surrounded by a disc. If both stars are disc-surrounded, disc material 
can be captured from the passing star, but the angular momentum of this captured matter is actually
extremely low \citep{pfalzner:apj05}. So including this captured matter would probably result in an 
even lower specific angular momentum. In this way the disc can be replenished with matter while
at the same time lowering its average specific angular momentum.
However, this increase in accretion can not become a runaway process as in this case 
accretion is directly coupled to disc mass loss, so this sets a natural 
limit on the material that can be accreted.

In the simulations herein, the starting point is a set of more or less fully-formed stars
describing the accretion process in the late stages of stellar formation. However, a similar 
mechanism could be at work earlier on in the star formation process, too. There 
the dominate process for the formation of massive stars is still an open question. 
Competitive accretion \citep{bates:06}, mergers \citep{bonnell:98,zinnecker:05} 
and accretion-easing processes \citep{krumholz:05} are the main mechanisms currently discussed. 
The present study suggests that {\it cluster-assisted accretion} should also be considered as a
formation process for massive stars. 

\section*{Acknowledgments}
Simulations were partly performed at the John von 
Neumann Institute for Computing, Research Center J\"ulich, Project HKU14.

\bibliographystyle{apj}




\end{document}